\documentclass{ws-mpla}
\usepackage[super]{cite}
\usepackage{graphicx,url,amsmath,amssymb}
\begin{document}

\markboth{J. I. Katz}
{Fast Radio Bursts---A Brief Review}

\catchline{}{}{}{}{}

\title{FAST RADIO BURSTS---A BRIEF REVIEW \\ SOME QUESTIONS, FEWER ANSWERS}

\author{\footnotesize J. I. KATZ}

\address{Department of Physics and McDonnell Center for the Space Sciences,
Washington University\\ St. Louis, Mo. 63130 USA \\ katz@wuphys.wustl.edu}

\maketitle

\pub{Received (Day Month Year)}{Revised (Day Month Year)}

\begin{abstract}
Fast radio bursts (FRBs) are millisecond bursts of radio radiation at
frequencies of about 1 GHz, recently discovered in pulsar surveys.  They
have not yet been definitively identified with any other astronomical object
or phenomenon.  The bursts are strongly dispersed, indicating passage
through a high column density of low density plasma.  The most economical
interpretation is that this is the interglactic medium, indicating that FRB
are at ``cosmological'' distances with redshifts in the range 0.3--1.3.
Their inferred brightness temperatures are as high as $10^{37\,\circ}$K,
implying coherent emission by ``bunched'' charges, as in radio pulsars.  I
review the astronomical sites, objects and emission processes that have been
proposed as the origin of FRB, with particular attention to soft gamma
repeaters (SGRs) and giant pulsar pulses.

\end{abstract}

\ccode{96.60.Gb; 97.60.Jd, 97.90.+j; 98.70.Dk}

\section{Introduction}
Fast radio bursts (FRBs), initially reported in 2007 \cite{L07}, were the
first major unexpected astronomical discovery in decades.  Although the
astronomical community was skeptical, concerned that the single ``Lorimer''
burst initially reported might have been interference, the discovery of four
additional bursts \cite{T13} removed most doubt and led to a surge of
research, both observational and theoretical.

Many FRB are so short ($\lesssim 1$ ms) that they are not temporally
resolved by the receivers, and their nominal measured burst lengths must be
considered only upper bounds on their intrinsic durations.  They are
therefore characterized by the time-integral of the flux $F_\nu$ received,
called the fluence:
\begin{equation}
{\cal F}_\nu \equiv \int F_\nu(t)\,dt,
\end{equation}
with units Jy\,ms\footnote{$1\,\mathrm{Jy} \equiv 10^{-26}\,
\mathrm{W/m^2\,Hz} = 10^{-23}\,\mathrm{erg/cm^2\,s\,Hz}$.}.

The most recent FRB Catalogue \cite{FRBcat} lists 17 bursts.  Since then,
one burst has been reported to repeat \cite{S16a,S16b}, with 17 sub-bursts
recorded over more than two years, eight of them clustered in a little over
an hour and four in about 20 minutes (because of the exigencies of observing
schedules, the absence of recorded bursts over a period of years does not
imply that the source was inactive during that time).  These sub-bursts
extend in fluence down to the detection threshold, and it is plausible that
more sensitive observations would detect many more.

Only one FRB has been associated with an astronomical object observed in any
other manner \cite{Ke16}, a galaxy at redshift $z = 0.492$ based on an
apparent radio flare from that galaxy lasting several days following the
FRB.  The FRB itself was only located within a Parkes beam approximately
$15^\prime$ across.  The statistical significance of this association is
controversial, and critics have suggested it is either an accidental
coincidence with a variable active galactic nucleus or the effect of
interstellar scintillation on a steady background source
\cite{WB16,V16,LZ16,AJ16}.  The ``afterglow'' (the prolonged
flare) had a fluence nearly $10^5$ times that of the FRB itself, which
argues against, but does not disprove, the reality of an association.  

Most known FRB were discovered by the High Time Resolution Universe project
at the Parkes radio telescope in Australia.  Their observed durations have
been in the range 1--10 ms.  The longer durations have been
frequency-dependent, varying roughly $\propto \nu^{-4}$, consistent with
multi-path broadening during their propagation \cite{W72}.  More
sophisticated theories \cite{R90} predict an exponent of $-4.4$, but
observations of pulsars indicate exponents scattered over a broad range,
mostly between $-3$ and $-4$ \cite{LDKK13,KMNJM15}.  Most measured fluences
are in the range $S = 1$--10 Jy-ms.  The lower end of this range is
approximately the limiting sensitivity for confident (signal to noise ratio
$\ge 10$) detection at Parkes.

The energy of an FRB doesn't arrive all at once.  Instead, the higher
frequency radiation arrives earlier, with a delay proportional to
$\nu^\alpha$, where $\nu$ is the radiation frequency and $\alpha = -2$ to
high accuracy.  This dispersion is the result of propagation through dilute
plasma, and its magnitude is proportional to the dispersion measure (DM),
the integral along the propagation path (to high accuracy, a straight line
of sight from source S to observer O) of the electron density $n_e$:
\begin{equation}
\mathrm{DM} \equiv \int_\mathrm{S}^\mathrm{O} n_e\,d\ell.
\end{equation}
Fitting DM to the raw measurement of flux {\it vs.\/} frequency is the first
step in FRB signal processing, as it has been in pulsar astronomy since
their discovery in 1967.  Then, subtracting the frequency-dependent time
delay, the fluxes across the spectrum are combined to give a
frequency-integrated pulse profile.

Radio telescopes are orders of magnitude more sensitive than detectors of
any other sort of radiation because of their large collecting areas (about
3000 m$^2$ for Parkes, 70,000 m$^2$ for Arecibo and nearly 200,000 m$^2$ for
the Chinese FAST now under construction, although the effective areas of 
fixed dishes are significantly less than their nominal areas, and depend on
the zenith angle) and because quantum noise is
negligible at radio frequencies.  Detection of weak radio sources is
possible only because of this sensitivity, but it has the consequence that
what is detected may be only an epiphenomenon representing a tiny fraction
of a source's total power.  For example, the radio emission of a pulsar may
be only $10^{-8}$ of its energy output.  This deprives the theorist of the
use of energy considerations as a tool for evaluating models.  
\section{Where Are They?}
The first question an astronomer asks about a newly discovered object is
its distance (its direction is usually known immediately from the instrument
used to detect it).  This is often a difficult question.  Distances are
usually impossible to determine directly because most astronomical objects
are too distant for trigonometric parallax, particularly if they are
observed only outside the visible band because the angular resolution of
radio and X-ray telescopes is very crude (parallax measurements also require
that the object remain detectable over months as the Earth moves around its
orbit).  Until the distance is determined, basic parameters such as
luminosity and brightness are unknown, often to many orders of magnitude,
and modeling is uncertain.  For example, the distances of quasars and active
galactic nuclei were controversial, even in order of magnitude, for several
years after their discovery, and the distances of gamma-ray bursts for
decades, and their theory was unguided until this uncertainty was removed.  

The DM conveys valuable information about the location of
a burst or pulsed source.  Within our Galaxy, it gives an estimate of the
path length of interstellar plasma through which the signal has propagated,
and hence the source's distance.  FRB have values of DM too large to be
explained in this manner, typically by a factor $\sim 10$, particularly
because most FRB are observed in directions far from the slab-like Galactic
disc, (at high ``galactic latitude'', with the Galactic disc defining an
equator).  After subtraction of the (small) estimated Galactic component,
the remaining DM must have another origin.  An analgous argument \cite{XH15}
shows that host galaxies, if they resemble our Galaxy, also cannot explain
the dispersion measures, except in cases of fortuitous edge-on lines of
sight.

There are two plausible sources of this extra-Galactic dispersion.  One is
the generic category of near-source plasma, whose nature depends on the
astronomical object and environment that make the FRB.  The other is the
dilute (mean density $\approx 1.6 \times 10^{-7} (1+z)^3$ cm$^{-3}$, where
$z$ is the cosmic redshift) intergalactic plasma.  No alternatives to these
two have been found to be plausible.

A host galaxy resembling our Galaxy could not provide most of the dispersion
measure unless it were fortuitously oriented almost exactly edge-on or the
FRB were very close to a dense plasma cloud at its center \cite{PC15}.
Clumping of intergalactic plasma into clouds, such as the intra-cluster
gas of clusters of galaxies, would not increase its mean density over its
cosmologically determined value.  Hence this could not produce mean DM
greater than that of a mean intergalactic medium, and cannot provide an
alternative to the inference of cosmological distances.

Determining whether the plasma dispersion is local to the source or results
from passage through the intergalactic medium is necessary to understanding
FRB because it determines their distances, astronomical environments, and
energy scale.  In this section I discuss the two possible hypotheses:
near-source and intergalactic plasmas.
\subsection{Near-source dispersion}
If the dispersion is produced by plasma close to and associated with their
sources, rather than by the intergalactic medium, then FRB may be our
neighbors, on cosmic scales, possibly even within our Galaxy.  At the high
Galactic latitudes of most FRB, a Galactic origin would suggest distances of
${\cal O}(100)$ pc ($1 \mathrm{pc} = 3.09 \times 10^{18}$ cm), the thickness
of the Galactic disc.  Greater ($\gtrsim 100$ kpc) distances $D$ have been
inferred from arguments based on the radiation emission of such plasma
clouds \cite{Ku14} and the spectral energy distribution of one FRB
\cite{Ku15}.  These distances would still be cosmologically local ($z \ll
1$; $D \ll 3$ Gpc).

Any hypothetical dispersing cloud local to the source must satisfy a
constraint on the electron density \cite{T14,D14}:
\begin{equation}
\label{nemax1}
n_e < {2 \over 3} \left\vert -\alpha -2 \right\vert {m_e \omega^2 \over
4 \pi e^2} \approx 5 \times 10^7\ \mathrm{cm}^{-3},
\end{equation}
where $\alpha \approx -2$ is the exponent in the dispersion delay $\delta t
\propto \nu^{\alpha}$ and $\omega$ is the (angular) frequency of the
observed radiation; the tightest observational upper bound on $\vert -\alpha
-2 \vert$ is 0.003, and other bounds are at most a few times larger; no
inconsistency with $\alpha = -2$ has been observed.

The observed DM of FRB are in the range 350--1600 pc\,cm$^{-3}$, using
astronomically convenient units, indicating dispersing clouds of dimension
$\gtrsim 3 \times 10^{13}$ cm, several hundred times the Solar radius.  This
bound is significant for models \cite{LSM14,ML15} involving dispersion in
stellar coron\ae\ or winds or other stellar phenomena.  These values of DM
are also far in excess of those of interstellar clouds, or of almost all
paths in the Galactic interstellar medium, the exceptions being paths that
pass very close to the Galactic nucleus \cite{PC15} or are almost exactly
aligned with the Galactic plane (only one of the 17 known FRB).  Similar
constraints apply to dispersion attributed to interstellar matter in a host
galaxy of a FRB.

A dispersing cloud must also be transparent to the observed radiation, which
gives an analogous, but temperature-dependent, upper bound on $n_e$
\cite{S62,LG14}:
\begin{equation}
n_e < 5 \times 10^3 {T_{8000}^{3/2} \over \mathrm{DM}_{1000}}\ 
\mathrm{cm}^{-3},
\end{equation}
where $T_{8000} \equiv T/8000^{\,\circ}$K and $\mathrm{DM}_{1000} \equiv
\mathrm{DM}/1000\,\text{pc\,cm}^{-3}$ (typical temperatures of dilute
ionized interstellar matter, set by a balance between photoionization
heating and radiative cooling, are $\sim 8000^{\,\circ}$K).  This bound
appears to be much stricter than that of Eq.~\ref{nemax1}, but in regions of
high energy density temperatures far in excess of $8000^{\,\circ}$K are
possible; in the Solar corona $T \sim 10^{6\,\circ}$K.

If the dispersion is to be attributed to something other than the
intergalactic medium, a origin associated with the FRB source itself must
be found.  One possibility is a region of dense plasma associated with a
galactic nucleus, such as that known (from observations of pulsar
dispersion) to exist within about 0.1 pc of the black hole (Sgr A$^*$) at
our Galactic center \cite{PC15}.  It would then remain to be explained why
FRB sources are invariably (every one of the 17 known) associated with such
plasma clouds or galactic nuclei; the implied plasma electron density of
${\cal O}(10^4\,\mathrm{cm}^{-3})$ bears no obvious relation to any proposed
FRB mechanism.

\begin{figure}
\centering
\includegraphics[width=3in]{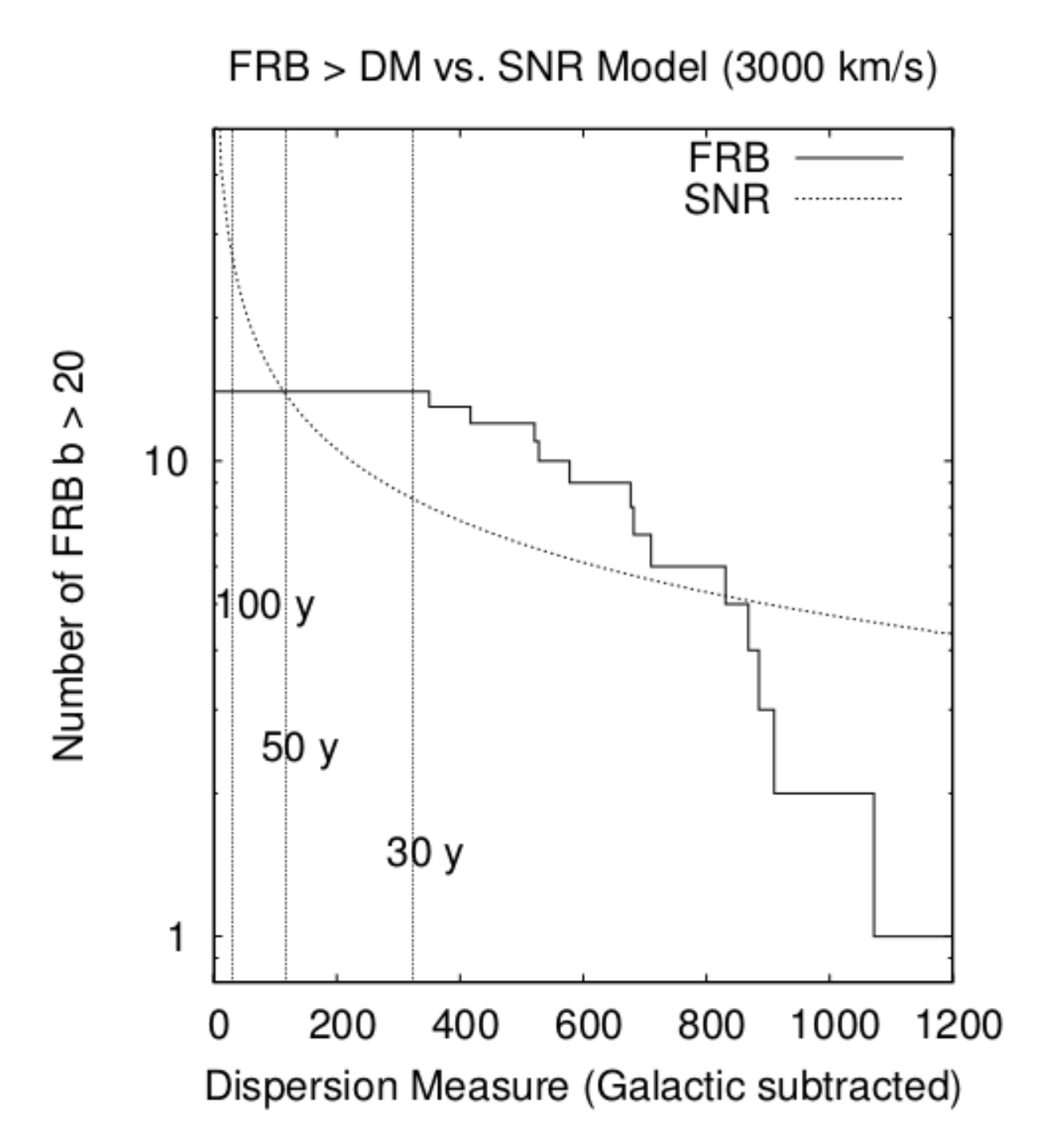}
\caption{\label{frbSNR} Observed (solid line) cumulative distribution of FRB
DM, in pc\,cm$^{-3}$, compared to fitted theory \cite{K16a} (dashed line) if
dispersion is produced by expanding supernova remnant shells.  Vertical
(dotted) lines show cutoffs in the predictions if FRB activity ceases at the
indicated ages, for a shell expansion speed $v = 3000$ km/s.  These cutoff
ages scale $\propto M_{shell}^{1/2}/v$.  Data from \cite{FRBcat}.}
\end{figure}

If FRB are found at the center of young supernova remnants (SNR), expanding
massive shells of gas expelled in a visible supernova, most of the
dispersion of the FRB may occur as it passes through the expanding plasma
shell of the SNR.  This hypothesis might explain the origin of the
dispersion, associates it with the formation of a compact object capable of
sudden energy release, like that of a FRB, and may be tested statistically.
Fig.~\ref{frbSNR} shows the predicted \cite{K16a} cumulative distribution of
FRB dispersion measures for high-Galactic latitude ($b > 20^\circ$) FRB,
after subtraction of the estimated Galactic contributions (generally less
than 10\% of the measured values); low $b$ FRB are excluded because for them
the Galactic contributions to DM are large and uncertain.

The dashed line is the predicted distribution with a fitted amplitude for a
SNR shell of one Solar mass ($2 \times 10^{33}$ g).  The vertical lines show
cutoffs if FRB sources turn off abruptly at the ages indicated, avoiding the
divergence of numbers at low DM, for an assumed expansion velocity of 3000
km/s; FRB searches may also exclude low-DM events in order to discriminate
against terrestrial interference.  Despite these caveats, the shape of the
distribution is clearly far from the prediction, evidence against SNR shells
as the origin of FRB dispersion.  No other model in which FRB are
cosmologically local has yet offered a plausible explanation or testable
predictions of the distribution of DM.

\subsection{Intergalactic dispersion}
If the dispersion of FRB is intergalactic, they are very distant, luminous
and bright.  Making standard cosmological assumptions \cite{I03,I04,DZ14},
the distance to a FRB may be inferred from its DM.  The results have been
$z$ in the range 0.3--1.3, and distances between 1 and 4 Gpc.  These
``cosmological'' distances imply, assuming isotropic emission, energies of
$\sim 10^{38}$--$10^{40}$ ergs for FRB.

\begin{figure}
\centering
\includegraphics[width=3in]{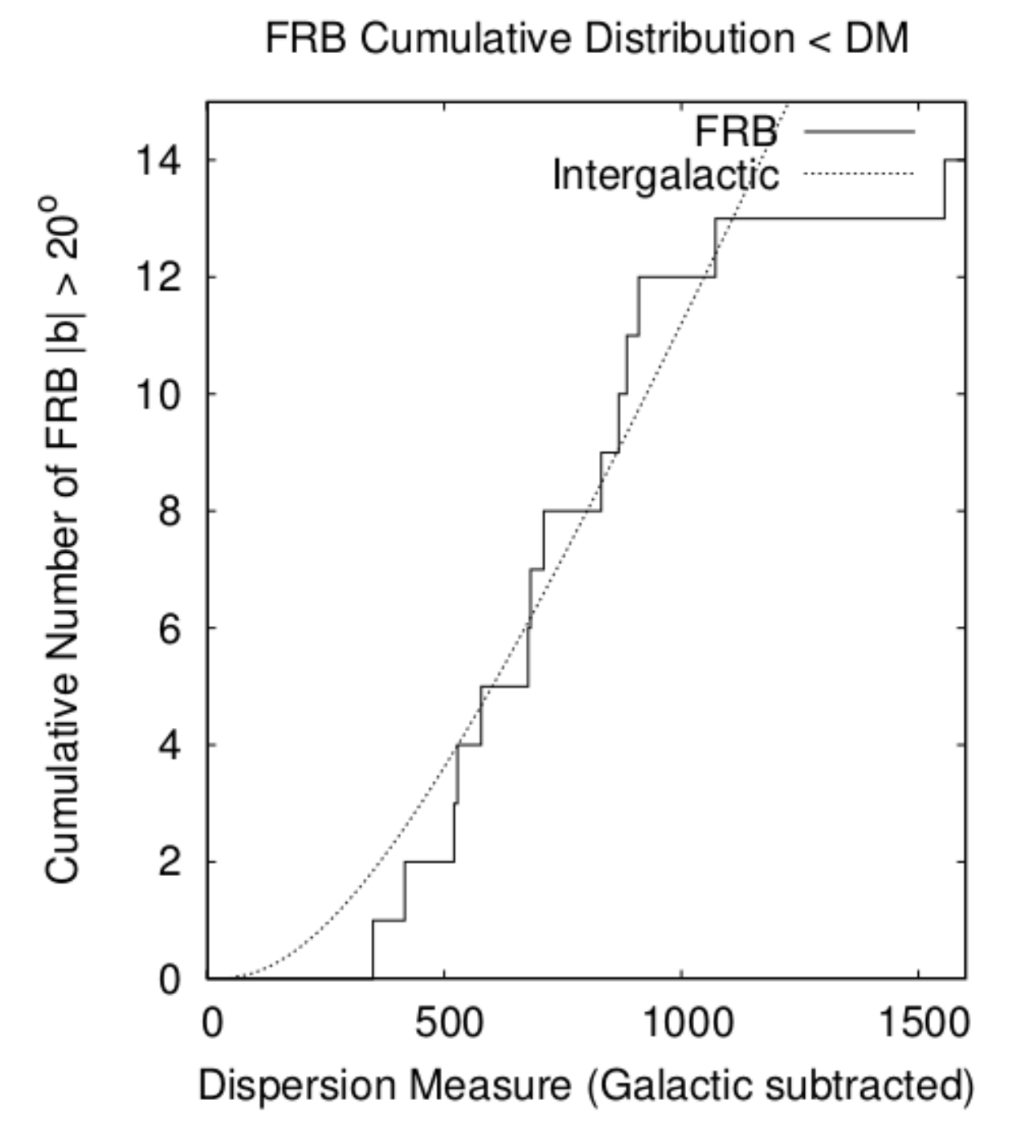}
\caption{\label{frbcosmo} Cumulative distribution of dispersion measures of
high Galactic latitude ($b > 20^\circ$) FRB.  Dashed curve is a simple model
of dispersion in the intergalactic medium, assuming a uniform detectable
source density, with fitted amplitude \cite{K16a}.  The data confirm the
predicted scarcity of low-DM FRB; the absence of FRB with high DM may be
attributed to their intrinsic faintness, to evolution of the event rate or
to reduced sensitivity of searches for brief transients at high DM.  Data
from \cite{FRBcat}.}
\end{figure}

This hypothesis predicts the distribution of DMs \cite{K16a}.
Fig.~\ref{frbcosmo} compares the predicted cumulative distribution for FRB
with $|b| > 20^\circ$ (because the Galactic contribution is large and
uncertain at lower $|b|$ only high-$|b|$ FRB are considered) to that
observed.  The most important predicted feature is the rarity of FRB with
small DM, a simple consequence of the Euclidean geometry of space in the
local Universe.  This is confirmed by the data (although searches may
exclude FRB with $\mathrm{DM} < 200$ pc\,cm$^{-3}$ to avoid terrestrial
interference).  The absence of FRB with very large DM may be attributed
either to their greater distances and redshifts (making them undetectably
faint), to cosmic evolution of the event rate or to reduced sensitivity of
their detection.

\begin{figure}
\centering
\includegraphics[width=3in]{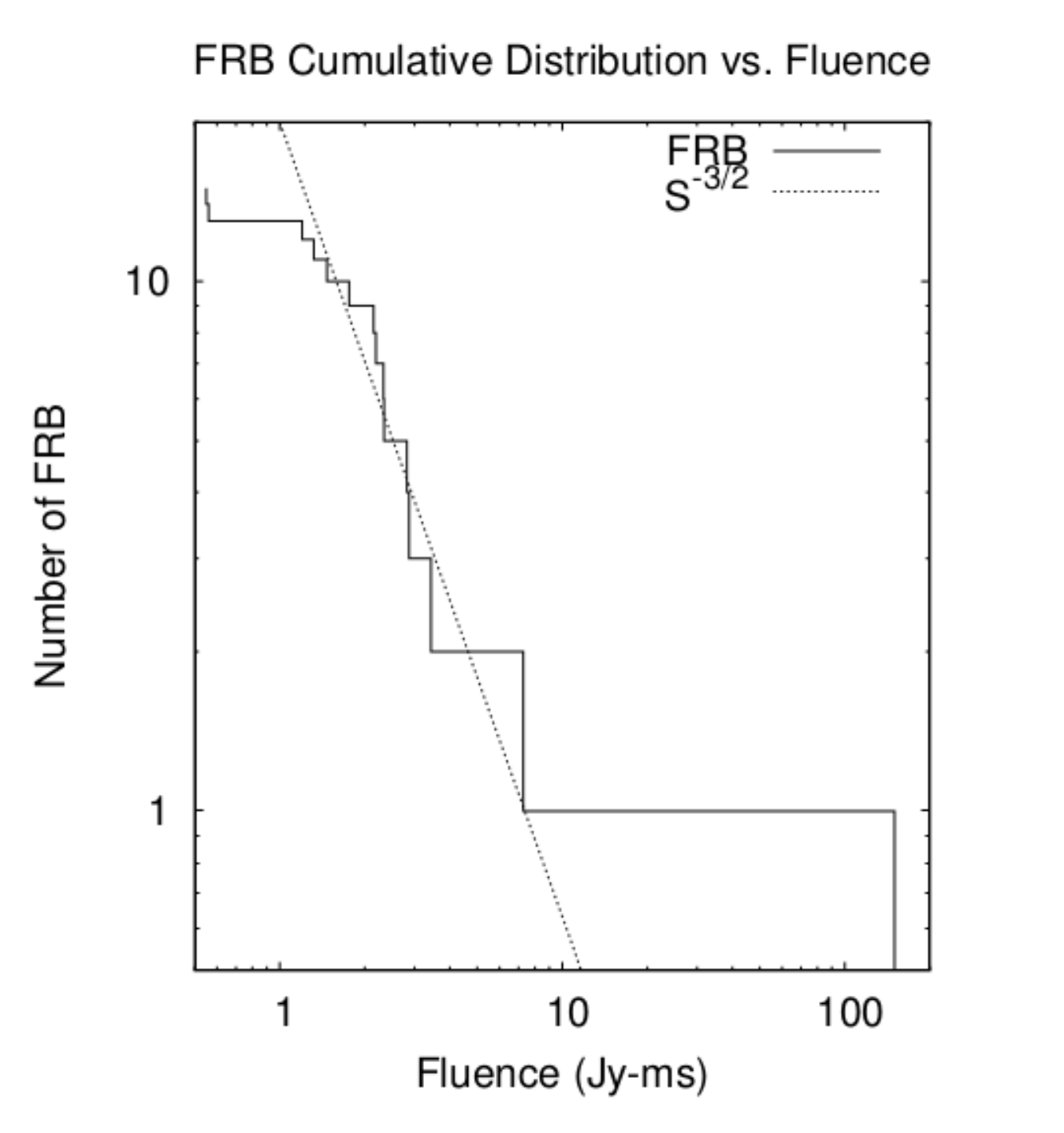}
\caption{\label{frbNvS} Distribution of cumulative number of observed FRB
{\it vs.\/} fluence, and a fitted $N \propto S^{-3/2}$ line, showing good
agreement.  No Galactic latitude cut is made because the interstellar medium
does not affect fluence measurements.  To maintain a homogeneous sample,
only the 15 FRB detected at Parkes are included; the single (repeating)
burst discovered at Arecibo and the single burst discovered by the Green
Bank Telescope are excluded.  However, the location of the observed FRB
within the Parkes beams is not known (with the possible exception of the
Lorimer burst, detected in three beams, from whose signal ratios a location
may be inferred), so all fluence values are given as if the
sources were at the centers of the detecting beams, and are in fact lower
bounds.  For sources randomly distributed on the sky this does not affect
the predicted exponent in the limit of large $N$.  Data from the FRB
Catalogue \cite{FRBcat} except for FRB 130628 \cite{C15}.}
\end{figure}

The hypothesis of cosmological distances makes an additional prediction.
For low redshifts, space is nearly Euclidean, the inverse square law applies
and the source function is close to its local value (because any cosmic
evolution may be expanded as a Taylor series in $z$).  Then the cumulative
number of sources $N \propto S^{-3/2}$, where $S$ may be any quantity that
follows an inverse square law.  For steady or slowly varying radio sources,
$S$ is the flux, and the failure of the $N \propto S^{-3/2}$ law
demonstrated the existence of cosmic evolution (and hence of the Big Bang).
For FRB, for which the peak flux cannot be deconvolved from the
effects of multipath propagation and instrumental response, we take $S$ to
be the burst fluence.  The results are shown in Fig.~\ref{frbNvS}.

They are consistent with the hypothesis of a uniformly filled Euclidean
space, except for the (previously noted \cite{L07}) anomalously bright
``Lorimer'' burst.  Perhaps this is accounted for by observational
selection: it is natural to impose extraordinarily conservative criteria
before accepting the reality of a new phenomenon.   A speculative
alternative is a spatially limited local enhancement of the burst rate.  The
deficiency of bursts with $S < 1$ Jy\,ms is likely attributable to the
detection threshold.

\begin{figure}
\centering
\includegraphics[width=3in]{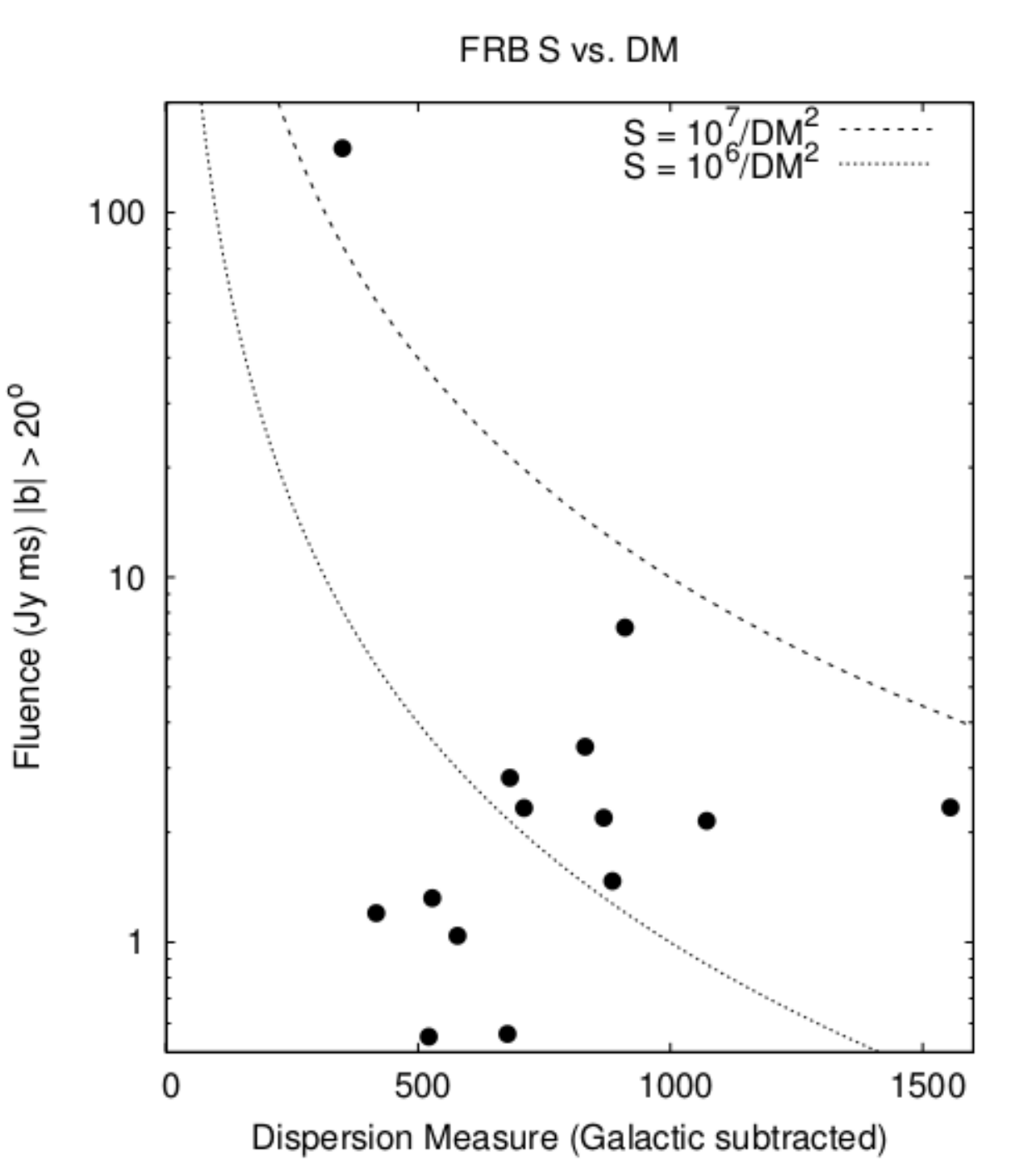}
\caption{\label{frbSDM} Distribution of FRB in $S$-DM space.  There is no
evidence that FRB are standard candles, and the apparent deficiency of weak
highly dispersed bursts may be a result of observational selection.  Data
from \cite{FRBcat}.}
\end{figure}

We also plot the fluence $S$ against DM in
Fig.~\ref{frbSDM}.  If the sources were standard candles in a Euclidean
universe they would lie on a curved line $S \propto \mathrm{DM}^{-2}$.  Two
curves of this form are shown.  Although Euclidean geometry is a good
approximation at these DMs, assuming the intergalactic
medium is the source of the dispersion, there is no indication that the
sources are standard candles.  The apparent deficiency of detections in
the lower right corner of the plot may be attributed to the difficulty of
detecting highly dispersed weak bursts.

An additional and independent argument in favor of ``cosmological''
distances of FRB is the recent discovery \cite{M15} of linear polarization
and measurement of Faraday rotation, parametrized by the rotation measure
RM, in FRB 110523.  Combining this with the DM gives the
electron density-weighted mean parallel component of magnetic field along
the propagation path
\begin{equation}
\langle B_\parallel \rangle_{n_e} = {\mathrm{RM} \over \mathrm{DM}} \equiv
{\int n_e B_\parallel\,d\ell \over \int n_e\,d\ell} = 0.37\ \mu\mathrm{G}.
\end{equation}
This is an order of magnitude less than typical interstellar magnetic fields
in our Galaxy and several orders of magnitude less than plausible fields in
denser plasma clouds.  It indicates that nearly all the dispersion occurs in
regions of very low (sub-$\mu$G) fields.  The only such regions known are
the intergalactic plasma and the gas within clusters of galaxies, but the
latter do not have sufficient dispersion measure to account for the observed
values.
\subsection{Pulse Broadening}
\begin{figure}
\centering
\includegraphics[width=3in]{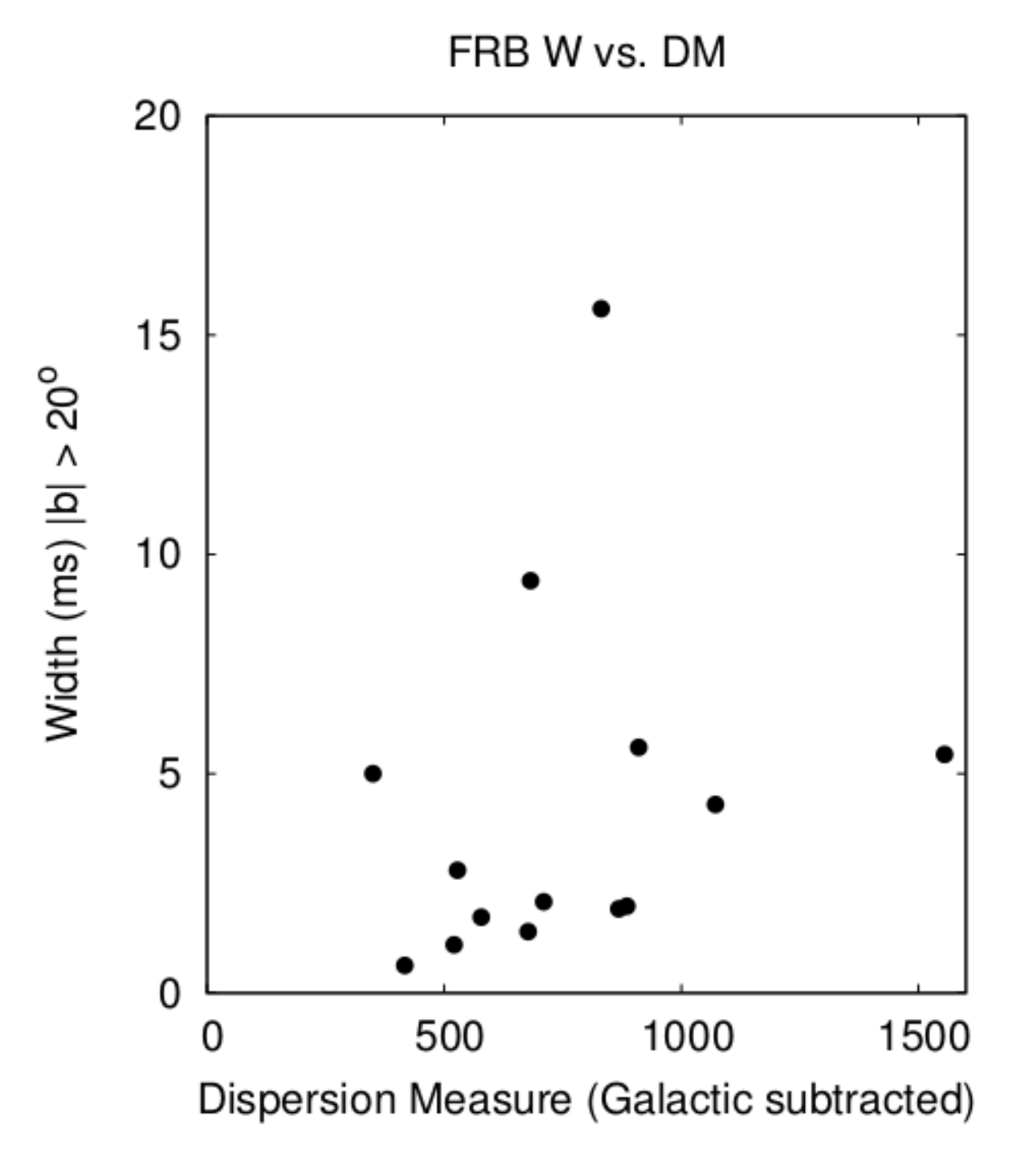}
\caption{\label{frbWDM} Propagation broadening widths (at 1400 MHz) of
high-latitude FRB.  The absence of correlation indicates the scattering does
not occur in the intergalactic medium.  Data from \cite{FRBcat}.}
\end{figure}

Several FRB have pulse widths $W$ greater than the resolution (after
de-dispersion) of the measurements, which is about 1 ms.  The steep
frequency dependence of $W$ (approximately $\propto \nu^{-4}$) indicates
that these widths result from multi-path propagation delays.  Plotting $W$
against DM (Fig.~\ref{frbWDM}) shows no correlation between these variables.
This implies that scattering in the intergalactic medium cannot be the cause
of the broadening, in agreement with theoretical arguments \cite{LG14}.

In several FRB $W$ is much greater than plausible for
Galactic propagation paths at their high Galactic latitude, and hence must
attributed to near-source regions.  However, the measured $W$ are not
atypical of those of Galactic pulsars with similar DM \cite{KMNJM15}, after
scaling to the frequencies at which FRB are observed, suggesting that
comparable regions may be found in our Galaxy.  These highly scattering
regions may be dense star-forming clouds; it has been suggested \cite{PC15}
that they may be the environs of galactic nuclei.

\section{What Are They?}
The fact that a FRB has been observed to repeat \cite{S16a,S16b} rules out
models that involve the destruction or irreversible transformation of the
source.  These include proposals of stellar collapse, merging binaries and
catastrophic collisions.

FRB are plausibly produced by compact young remnants of stellar collapse,
neutron stars or black holes, whose deep gravitational potential wells 
permit the sudden emission of energy.  This hypothesis has been incarnated
as soft gamma repeaters (SGRs) \cite{Ku14,Ku15,PC15,K16a,PP07,PP13,L14,K15}
or giant pulses from young pulsars \cite{K12,K16a,CW15,CSP15,LBP16}; these
are all rare, episodic but repeating events of low duty factor.  Related
ideas have included the interaction of pulsars with planets \cite{MZ14},
asteroids or comets \cite{GH15,DWWH16}.  Less exotic models, applicable only
if FRB are not at cosmological distances, have appealed to events like
stellar flares \cite{LSM14,ML15}.

The observed short durations of FRB imply small emitting regions, because
emission over larger regions would, by a spread in radiation travel times
$\Delta t$, produce longer pulses; in the absence of relativistic bulk 
(including phase) motion $\Delta t \ge \Delta r/c$, where $\Delta t$ is the
observed burst duration and $\Delta r$ is the dimension of the radiating
region (properly, its dimension along the direction of radiation).

The observed $\Delta t \lesssim 1$ ms (before broadening by multipath
propagation) implies small source regions, and a correspondingly high
radiation intensity at the source.  This is usually described by a
 ``brightness temperature'' $T_b$:
\begin{equation}
T_b \equiv {F_\nu c^2 \over 2 \nu^2 k_B},
\end{equation}
the temperature (in the Wein limit, applicable at radio frequencies) of a
black body emitting the flux density of radiation $F_\nu$, whose units are
ergs/s\,cm$^{2}$\,Hz\,sterad.  Because the angular sizes of compact sources
are not directly measured, it is usually assumed that their emitting areas
$A \sim (c \Delta t)^2$ and that $F_\nu \sim F_{\nu,obs}(4 \pi D^2/A)$,
where $F_{\nu,obs}$ is the observed spectral density integrated over the
(unknown) source solid angle, and has units ergs/s\,cm$^2$\,Hz, tacitly
assuming isotropic emission.  By Liouville's theorem, the brightness
temperature of the radiation we observe is the same as that at the source;
we just observe it in the tiny (not directly observed, but calculated from
the inferred source size and distance) solid angle subtended by the source,
while at the source it is assumed to fill steradians.

FRB have $T_b$ up to $\sim 10^{37\,\circ}$K.  Of course, these extraordinary
values do not indicate any emitter with that physical temperature, or even
particles with the corresponding energy $k_B T_b$.  Just as for pulsars,
extraordinary $T_b$ indicate coherent radiation by ``bunches'' or coherent
waves containing large numbers of particles \cite{K14}.  In fact, the
``nanoshots'' of at least one pulsar had $T_b$ higher \cite{S04} even than
those of FRB.  Theoretical plasma physics has so far been incapable to
explaining the high brightness of radio pulsars (without which they would
be unobservably faint), and this is likely to be the most difficult aspect
of FRB to understand.

The short durations of FRB (the upper limit on the light travel times across
their sources ($c \Delta t < 3 \times 10^7$ cm) exclude a source region
larger than a neutron star (radius $R = 10^6$ cm) or stellar mass black hole
(Schwarzschild radius $3 \times 10^5 (M/M_\odot)$ cm, where $M_\odot$ is the
Solar mass).  They permit small emitting subregions of larger objects, such
as stellar flares \cite{LSM14,ML15}.  The only other astronomical phenomena
with time scales as short as those of FRB are the rise times of the giant
flares of Soft Gamma Repeaters, which have been observed to be 200--300
$\mu$s \cite{K16a}, and pulses, subpulses and ``nanoshots'' of radio pulsars
\cite{HKWE03,S04,HE07} some of which have durations $\le 0.4$ ns.  All of
these are produced by neutron stars.  Neutron stars are also regions of high
gravitational, and in some cases high magnetic, energy density, and hence
natural origins of energetic events.  Two classes of sudden neutron star
outbursts have been considered candidates for the origin of FRB: 
\subsection{Soft Gamma Repeaters}
\label{SGR}
SGR were suggested as candidate FRB sources shortly after their discovery,
and have been advocated many times since
\cite{Ku14,Ku15,PC15,K16a,PP07,PP13,L14,K15}.  SGR and FRB have several
similarities: characteristic time scales, low duty factors and repetition.
They also have an important difference: SGR appear to be entirely thermal
phenomena, radiating black body-like spectra of X- and gamma-rays.
Heterogeneous temperatures mean that the integrated spectra need not be
Planckian, but they are heavily self-absorbed at low frequencies, with
brightness temperatures close to the material temperature; SGR outbursts are
not observed at frequencies below the X-ray range.  This is entirely unlike
FRB, which are observed only in radio waves, with extraordinarily high
brightness temperatures.

Attempts to explain FRB as a consequence of SGR sources ({\it N.B.:\/} they
might not be associated with SGR outbursts themselves, even if produced by
the same strongly magnetized neutron stars) have had to appeal to models of
``magnetar'' (hyper-magnetic) neutron star magnetospheres with $B \sim
10^{14}$--$10^{15}$ G \cite{K82,TD92,TD95,BT07}.  The model assumes that a
substantial part of the neutron star's magnetic moment, inferred from its
spin-down rate by treating it as a rotating dipole in vacuum, has its source
in currents flowing through the near-vacuum magnetosphere rather than the
dense neutron star interior.  In magnetar models of SGR these currents may
be induced by fracture and motion in the neutron star's solid crust, but in
the FRB model \cite{K15} they were frozen-in during the collapse that formed
the neutron star.  The currents cannot be rapidly interrupted because the
large circuit inductance would produce an electromotive force that would
spark across the gap; instead, they decay slowly, over hundreds to thousands
of years, consistent with the ages of the neutron stars identified as SGR
sources.

The ultimate energy source for both the SGR and FRB activity is the
magnetostatic energy of the neutron star.  Although a magnetic field
may be characterized by a magnetic energy density $B^2/8\pi$, this energy
is global rather than local and cannot be carved out like a scoop of ice
cream.  Even though the magnetostatic energy density may be high in vacuum,
tt cannot be released there because there is no charge on which an induced
electric field can do work.  If energy is released below the neutron star's
surface, it diffuses to the surface as thermal X-rays over an extended
period of time, explaining neither the rapid rise of a SGR outburst nor a
FRB.  Hence it must be released in the current-carrying magnetosphere, and
SGR and FRB activity stops when the magnetospheric currents decay.

To explain FRB it is necessary to assume that when the right conditions are
met the impedance along the magnetospheric current path suddenly increases,
leading to rapid dissipation of energy.  This is the classic mechanism of
magnetic reconnection \cite{PF00}.  In this model it may be produced by
single-particle Coulomb scattering because electrostatic quasi-neutrality
requires that the magnetospheric ion density be proportional to the current
density.  If, as a result of flow in the neutron star itself that changes
the frozen-in magnetic field, the magnetospheric current density increases,
the electron density must also increase (because the current-carrying
electrons are relativistic), and quasi-neutrality requires an equal increase
in the ion density.  But electron-ion scattering limits the mean electron
velocity to $c$ divided by the number of scatterings along a magnetospheric
path.  This imposes a maximum current density because the mean electron
velocity is inversely proportional to the ion density, and hence to the
electron density \cite{K15}.  This is in contrast to an ordinary plasma in
which the mean electron velocity can increase to carry an increasing
inductively-driven current density.

The result is likely to be a sudden increase in electromotive force and
energy deposition as the current fails to keep up with its inductive drive.
Then, by a {\it deus ex machina\/}, plasma instability leads to intense
coherent emission.  This plausibly occurs during the rapid rise of a SGR
giant flare, occurring on a sub-ms time scale is consistent with that of
FRB (even though the full width of a SGR flare, emitting a thermal spectrum,
is typically hundreds of ms), before the released energy has thermalized.
This hypothesis is difficult to evaluate or test, an unsatisfactory state of
affairs, also true of pulsar emission mechanisms.

Unfortunately, this model of FRB as counterparts of SGR may fail because the
Parkes telescope did not detect a FRB when it was fortuitously observing
during an outburst of SGR 1806-20 \cite{TKP16}.  The telescope was pointing
away from the SGR, but even its far side-lobe sensitivity is $\sim 10^{-6}$
of its main-lobe sensitivity, while a FRB at the distance of the SGR (about
9 kpc \cite{BCFC08}) would be expected to be about $10^{11}$ times as bright
as one at typical ``cosmological'' distances of $\sim 3$ Gpc.  The SGR model
of FRB predicts \cite{K14,K16a} that any radio telescope searching (with
high time resolution and de-dispersing signals) for transients or pulsars
would detect an extraordinarily strong FRB signal, equivalent to $\sim 10^5$
Jy\,ms ($10^4$--$10^5$ times the fluence of observed FRB) in-beam, during a
Galactic SGR outburst that is above its horizon.  The FRB would show the
dispersion of the Galactic interstellar medium between it and the observer;
the DM of SGR 1806-20 is not known because it is not observed at radio
frequencies, but it might be expected, based on its distance in the Galactic
plane, to be $\sim 300$ pc\,cm$^{-3}$.

There may be loopholes in this argument that the failure to detect a FRB
simultaneous with SGR 1806-20 excludes the SGR origin of FRB \cite{K15}.
Perhaps the relation between observable FRB and SGR is not 1:1.  For
example, the extremely nonthermal FRB emission may be strongly beamed.
Alternatively, a FRB may have been broadened in propagating through the
interstellar medium by an amount outside the 14--56 ms broadening window
used in the observations at the time of SGR 1806-20 \cite{TKP16}.
\subsection{Giant pulsar pulses and nanoshots}
Some radio pulsars emit, in addition to their regular pulses with durations
of ms (or tens or hundreds of $\mu$s for ``millisecond'' pulsars with
periods of 1--10 ms) very intense and much shorter bursts called
``nanoshots'' \cite{S04,HKWE03,HE07}.  Although these are not energetic
enough to be detected at distances of hundreds of kpc or greater, they are
empirical evidence (not theoretically understood) of pulsar
behavior.  Can analogous behavior on much higher energy scales produce FRB
even at cosmological distances?
\subsubsection{Nanoshots}
It is first necessary to consider the physics of nanoshots and its possible
extrapolation to higher energies.  Two very different pulsars have been
observed to produce nanoshots.  One was the Crab pulsar, with a spin period
of 33.5 ms and (polar, in a dipole model) magnetic field $B_p = 4 \times
10^{12}$ G.  The other was PSR B1937+21, a millisecond pulsar with a spin
period of 1.558 ms and (polar, in a dipole model) magnetic field $B_p = 4
\times 10^8$ G.   PSR B1937+21 is believed to be a ``recycled'' old pulsar,
whose magnetic field decayed and spin slowed, but that was subsequently
spun-up by accretion from a (now lost) close binary companion.  In contrast,
the Crab pulsar is the product of a supernova in the year 1054.  Taking the
distance to the Crab pulsar $D = 2.2$ kpc and $D \ge 3.6$ kpc to PSR
B1937+21, and assuming isotropic emission, their nanoshot energies and
durations were
\begin{equation} 
E = 4 \pi D^2 {\cal F}_\nu \Delta \nu = \begin{cases} 1.0 \times 10^{28}\
\mathrm{ergs} \quad \Delta t \le 0.4\ \mathrm{ns} &\text{Crab} \\
1.2 \times 10^{27}\ \mathrm{ergs} \quad \Delta t \le 15\ \mathrm{ns} 
&\text{B1937+21} \\ 1 \times 10^{40}\ \mathrm{ergs} \quad \Delta t \le
1\ \mathrm{ms} &\text{FRB}, \end{cases}
\end{equation}
where the fluence integrated over the bandwidth $\Delta \nu$ was measured
and corresponding numbers for FRB are included.

If the source regions were not moving relativistically, their volumes may be
estimated as $\sim (c \Delta t)^3$.  Supposing it were possible to
annihilate magnetostatic energy in the source region and to radiate it as
the radio-frequency energy of a nanoshot with unit efficiency, the lower
bound on the nominal corresponding magnetic field:
\begin{equation}
B_{nom} = \sqrt{8 \pi E \over (c \Delta t)^3} \gtrsim \begin{cases} 1.2
\times 10^{13}\ \mathrm{G} & \text{Crab} \\ 1.8 \times 10^{10}\ \mathrm{G} &
\text{B1937+21}.\\ 5 \times 10^{11}\ \mathrm{G} & \text{FRB}, \end{cases} 
\end{equation}
where for the FRB $c \Delta t$ is replaced by a neutron star radius $R =
10^6$ cm.  For the two pulsars, the implied magnetic fields and energy
densities appear to exceed those at the surfaces of the neutron stars by
large factors, a problem that was pointed out by the discoverers of
nanoshots \cite{S04,HE07}, indicating the failure of a na{\"i}ve model based
on magnetostatic energy.  It is conceivable that the local fields are much
larger than the dipole fields, as in sunspots; for the fast-rotating PSR
B1937+21 this hypothesis predicts a large, and perhaps measurable, braking
index if high magnetic multipole moments dominate the spindown.

Even if $B_{nom}$ were less than $B_p$, it would not explain the origin of
the nanoshot energy.  A vacuum magnetic field, whose source is currents
within the star, cannot dissipate energy.  But the pulsars known to emit
nanoshots are rapidly rotating, suggesting that rotation is essential (in
contrast to SGR/magnetars whose physics is believed to be that of a
non-rotating magnetosphere).  The spindown torque is exerted on the star
through currents flowing on its open field lines \cite{GJ69}.  They
intersect the stellar surface on a pole cap of area $\pi R^3 \Omega/c$,
where $\Omega$ the angular frequency of rotation.  The spindown power
density on the polar cap is
\begin{equation}
I = {1 \over 6} {(B_p R^3)^2 \Omega^4 \over c^3} {c \over \pi R^3 \Omega} =
{B_p^2 c \over 6 \pi} \left({\Omega R \over c}\right)^3.
\end{equation}
Some unknown portion of this may be available to power nanoshots.

If the plasma source of the nanoshots is moving towards the observer with a
bulk Lorentz factor $\Gamma$, then only a solid angle $\sim \Gamma^2$ sterad
is illuminated, rather than $4 \pi$ sterad, with a corresponding reduction
in the total energy required.  The duty factor of observed nanoshots is very
small, so this may be consistent with their observation in a significant
fraction, perhaps all (unsuccessful searches do not appear to have been
published), of the pulsars examined at high time resolution. 

Bulk relativistic motion also implies that the radius of the emitting
region that is observed over an interval $\Delta t$ is $\sim c \Delta t 
\Gamma$, with area $\sim (c \Delta t \Gamma)^2$, and its depth along the
line of sight is $\sim c \Delta t \Gamma^2$.  The volume that can contribute
to the nanoshot is $\sim (c \Delta t)^3 \Gamma^4$, implying a total energy
\begin{equation}
E_{max} \sim {B_p^2 c \over 6b\pi} \left({\Omega R \over c}\right)^3
(c \Delta t)^2 \Delta t \Gamma^4 \sim \begin{cases} 4 \times 10^{20}
\Gamma^4\ \mathrm{ergs} & \mathrm{Crab} \\ 2 \times 10^{21} \Gamma^4
\ \mathrm{ergs} & \text{B1937+21}. \end{cases}
\end{equation}
These values should be compared to the inferred nanoshot energies, allowing
for beaming,
\begin{equation}
E \sim \begin{cases} 1 \times 10^{27} \Gamma^{-2}\ \mathrm{ergs} &
\mathrm{Crab} \\ 1 \times 10^{26} \Gamma^{-2}\ \mathrm{ergs}&
\text{B1937+21}.
\end{cases}
\end{equation}
The maximum theoretical energies are consistent with observations if
$\Gamma \gtrsim 10$.  Of course, this does not explain how the spindown
power makes the observed nanoshots; it only shows that it is consistent
with the energetic constraints.
\subsubsection{FRB?}
These arguments can be modified to apply to giant pulse models of FRB.  Only
bounds on FRB intrinsic durations are known, and these bounds are $\sim
10^6$ times longer than the measured nanoshot durations.  The condition that
the FRB energy not exceed the product of spindown power and its intrinsic
duration
\begin{equation}
\label{Emax}
E < {1 \over 6} {(B_p R^3)^2 \Omega^4\over c^3} \Delta t = 6 \times 10^{46}
B_{15}^2 \Omega_4^4 \Delta t_{-3}\ \mathrm{ergs},
\end{equation}
where $B_{15} \equiv B_p/10^{15}\,\mathrm{G}$, $\Omega_4 \equiv \Omega/10^4
\,\mathrm{s}^{-1}$ and $\Delta t_{-3} \equiv \Delta t/10^{-3}\,\mathrm{s}$;
the dimensionless parameters have been normalized to their maximum credible
values.  Equivalently, the source parameters are constrained
\begin{equation}
\label{powerparameters}
B_{15}^2 \Omega_4^4 \Delta t_{-3} > 1.6 \times 10^{-7} E_{40},
\end{equation}
where $E_{40} \equiv E/10^{40}\,\mathrm{ergs}$.  There is ample room in 
parameter space to satisfy this inequality, even allowing for inefficiency
in converting rotational energy to radiation.  The bounds of Eqs.~\ref{Emax}
and \ref{powerparameters} use the actual pulse energy $E$, that may be less,
perhaps by a large factor, than that inferred from the measured fluence by
assuming isotropic emission.

The observation of repetitions of FRB 121102 over nearly three years
\cite{S16a,S16b} sets a lower bound on the spin-down time
\begin{equation}
\label{spindown}
t_\text{spin-down} = 3 {I c^3 \over B_p^2 R^6 \Omega^2} \approx {10^3 \over
B_{15}^2 \Omega_4^2} \mathrm{s} > 10^8\ \mathrm{s},
\end{equation}
where $I \approx 10^{45}$ g\,cm$^2$ is the neutron star moment of inertia,
or 
\begin{equation}
\label{BOmega}
B_{15}^2 \Omega_4^2 < 10^{-5}.
\end{equation}
Combining with Eq.~\ref{powerparameters} yields
\begin{equation}
\Omega_4^2 \Delta t_{-3} > 1.6 \times 10^{-2} E_{40}.
\end{equation}
Because $\Omega_4$ cannot much exceed unity, this sets a non-trivial lower
bound on $\Delta t$ of about $10^{-5}$ s in the model of FRB as giant pulsar
pulses at cosmological distances.  This bound is relaxed if the observed
emission is beamed towards us.

The observation that $\Delta t_{-3} \le 1$ sets a lower bound on the spin
frequency
\begin{equation}
\Omega_4 > 0.13 E_{40}^{1/2};
\end{equation}
the spin period of the most energetic (observed, and assuming isotropic
emission) FRB cannot exceed 5 ms.  Then Eq.~\ref{BOmega} implies 
\begin{equation}
B_{15} < {1 \over 300 \Omega_4} < {1 \over 40} \sqrt{\Delta t_{-3} \over
E_{40}}.
\end{equation}
Combining Eqs.~\ref{powerparameters} and \ref{spindown} bounds the spin-down
time
\begin{equation}
t_\text{spin-down} < 200 {\Delta t_{-3} \over E_{40}} \Omega_4^2\
\mathrm{y}.
\end{equation}
Again, this bound is relaxed if the burst is beamed.

If FRB are produced by such rapidly slowing neutron stars then they
must be quite young, and perhaps the product of supernov\ae\ in the era of
photographic, or even CCD, astronomy.  The contribution of the remnant, if
fully ionized and not clumped (contrary to expectation \cite{HS84}), to the
dispersion measure would be
\begin{equation}
\mathrm{DM}_\mathrm{SNR} = {M_\mathrm{SNR} \over M_\odot} {30\ \mathrm{pc}\,
\mathrm{cm}^{-3} \over (A_d v_{30,000})^2},
\end{equation}
where $A_d$ is the SNR's age in decades, $M$ its mass, $v_{30,000} \equiv
v/30,000\,\text{km/s}$ and $v$ is its expansion velocity.  The absence of
any significant change in the dispersion measure of the repeating FRB 121102
\cite{S16a,S16b} over nearly three years thus sets a lower bound, if the
remnant is spherically symmetric and ionized (for example, by collision with
surrounding gas or by internal shocks) $A_d v_{30,000} > 3$ in this model.

The preceding numerical estimates combine the shortest $\Delta t$, the
greatest $E$ and the span of repetitions of the sole FRB that has been
observed to repeat as if they described the same object, and assume
isotropic emission if $E$ is obtained from the observed fluence.  In fact,
they were observed for different FRB, so that the numerical inferences
tacitly assume that all FRB have similar characteristics, as well as Occam's
assumption of the simplest possible interpretation.

The hypothesis that FRB are giant pulsar pulses at cosmological distances
requires them to be in a fairly narrow corner of parameter space, but it is
not excluded.  It leads to a number of qualitative predictions, or at
least suggestions:
\begin{enumerate}
\item FRB are preceded by supernov\ae, probably by years to a century.
\item FRB dispersion measures will decrease to an asymptotic (intergalactic
medium) value if the supernova remnant makes a measurable contribution.
\item Repetitions of FRB will continue over their spin-down times of years
to a century.
\item The repeated bursts will gradually decrease in energy as the neutron
star slows and its spindown power decreases.
\item Even at cosmological distances, FRB may be observable as millisecond
pulsars, perhaps either at radio or visible frequencies, with spindown times
of years to a century (longer if beamed). 
\end{enumerate}
\subsubsection{Propagation}
Quite apart from their demonstration of intense emission, the observations
of nanoshots showed that nanosecond pulses can travel through the plasma of
the Galactic plane for substantial distances (about 2.2 kpc for the Crab
pulsar and $\ge 3.6$ kpc for PSR B1937+21) without significant broadening
by multipath propagation.  The nanoshots are less broadened than would be
expected on the basis of the broadening of pulsar pulses measured with
poorer temporal resolution at these distances and dispersion measures
\cite{LDKK13,KMNJM15}.

Perhaps there are, in fact, multiple propagation paths, with significant
time delays between them, so that in observations with nanosecond temporal
resolution the different paths produce distinct, apparently unrelated, spikes
rather than a smoothly broadened pulse.  In
geometrical optics a focus forms at an extremum of the optical path (travel
time), so that a single image will not be broadened if diffractive effects
are negligible.  That hypothesis would imply even more energetic nanoshots
because the detected energy would be only a fraction of the total.
\section{Discussion}
The development of multi-dish radio telescopes, planned to culminate in the
square kilometer array with collecting area of $10^6$ m$^2$ \cite{CR04},
more than an order of magnitude greater than that of Arecibo and five times
that of FAST, will lead to the observation of much larger numbers of fainter
FRB.  The faint (some as weak as 0.1 Jy\,ms) repetitions of FRB 121102 were
observed at Arecibo and the Green Bank Telescope, with collecting areas
22 and 2.4 times that of Parkes, respectively.  The actual detection rate
will depend on how the telescope is used (many simultaneous beams with
moderate sensitivity or a single beam collecting energy from its entire
aperture), on unknown properties of the FRB population: their intrinsic
``luminosity function'' (event rate as a function of radiated energy, more
properly called a fluence function because it is the fluence rather than the
flux that is measured), and their spectra and distribution in the Universe
because redshift affects the detectability of cosmologically distant events.

Positive identification of a FRB with some other astronomical object would
greatly advance our understanding.  The recognition of trains of repetitive
outbursts of a single FRB is a giant step in this direction because it will
permit the use of interferometry to determine an accurate position on the
sky (observations at a single telescope can only determine position to
approximately its beam width, about $15^\prime$ at Parkes).

Identification would likely immediately resolve the question of the distance
scale to FRB, because at least approximate estimates of the distances to
most other classes of astronomical objects are known, and measurement of 
their redshift gives an immediate and accurate distance measurement of
cosmologically distant objects.  The critical step to understanding of
gamma-ray bursts, after more than two decades of perplexity, was their
identification on the basis of temporal coincidence with visible light
transients whose coordinates were determined to arc-seconds by visible
imaging.   Then identification with galaxies with measurable redshifts was
immediate, and provided conclusive proof that they are at cosmological
distances.  The suggested FRB association with a distant galaxy \cite{Ke16}
would be equally conclusive, if it is confirmed, either by statistical
arguments or by another identification.  Identification might also give
clues as to the FRB mechanism if they are associated with some sort of
peculiar object.

The fluences of FRB have complex non-monotonic frequency
dependence \cite{T13}.  The several bursts of the repeating FRB 121102 show
that this changes from burst to burst \cite{S16a,S16b}.  Studies of the
frequency structure of the single burst FRB 110523 \cite{M15} indicate that
this structure may be attributable to scintillation produced by refraction
along the propagation path \cite{C90}.  This is consistent with the rapid
(on time scales of minutes) changes in spectrum of the repeating FRB.  From
this rapidly growing body of data, and comparison with the scintillation of
Galactic pulsars, it will be possible to extract information about the
plasma environments of FRB.

The reader will note the large number of references from 2016 in this
review that was completed March 31, 2016.  This subject is developing
rapidly.  We may hope that an understanding of the astronomical nature and
environments of FRB will soon be developed, even if the mechanism of their
coherent emission remains as enigmatic as that of pulsars, which they may
resemble.

\end{document}